# Getting Humans to do Quantum Optimization - User Acquisition, Engagement and Early Results from the Citizen Cyberscience Game Quantum Moves

ANDREAS LIEBEROTH, Aarhus University

MADS KOCK PEDERSEN, Aarhus University

ANDREEA CATALINA MARIN, Aarhus University

TILO PLANKE, Aarhus University

JACOB FRIIS SHERSON, Aarhus University

## ABSTRACT

The game Quantum Moves was designed to pit human players against computer algorithms, combining their solutions into hybrid optimization to control a scalable quantum computer. In this midstream report, we open our design process and describe the series of constitutive building stages going into a quantum physics citizen science game. We present our approach from designing a core gameplay around quantum simulations, to putting extra game elements in place in order to frame, structure, and motivate players' difficult path from curious visitors to competent science contributors. The player base is extremely diverse – for instance, two top players are a 40 year old female accountant and a male taxi driver. Among statistical predictors for retention and in-game high scores, the data from our first year suggest that people recruited based on real-world physics interest and via real-world events, but only with an intermediate science education, are more likely to become engaged and skilled contributors. Interestingly, female players tended to perform better than male players, even though men played more games per day. To understand this relationship, we explore the profiles of our top players in more depth. We discuss in-world and in-game performance factors departing in psychological theories of intrinsic and extrinsic motivation, and the implications for using real live humans to do hybrid optimization via initially simple, but ultimately very cognitively complex games.

## 1. INTRODUCTION

When online participants are used as workhorses for difficult problems such as EteRNA's needle-in-haystack-like RNA model selection task (Lee et al, 2014) or EyeWire's formidable challenge of mapping neural connectivity in the mouse retina (Marx, 2013), conclusive results may lie months and years into the future. In the rapidly growing field of human computation the design of new initiatives is often based on intuition rather than proven design hypotheses. The citizen cyberscience community is, however, a remarkably open scientific group, which allows us to capitalize on a unique and generous culture to learn from each other at each step of the journey – not just in the end, when all is securely tested and published.

The citizen cyberscience game Quantum Moves was designed to help build a quantum computer - a computer more powerful than any other in the world based on moving atoms around under the principles governing quantum physics. This delicate process involves a constant risk of losing hold on the volatile atoms, if they are not moved precisely and quickly. The simulations in which our algorithms tried to optimize this process bear a remarkable similarity to side scrolling casual games, and so the notion of human quantum optimization was hatched: What if real humans would do things differently than the logical step-by-step nature of the algorithms? Would explicit understandings of the counterintuitive quantum problems help people solve our problem more intelligently? Would human physical and cognitive fallibility add an interesting random factor? And even if people in general could not beat the AI, would play trajectories resulting from certain lucky punches or persistent quantum heroes be enough to help the AIs learn in new ways? This is the premise of human-computer hybrid optimization: Helping AIs learn through real people's blooming buzzing mess of solutions, when problems can be represented as engaging game levels.

Quantum Moves places itself with EyeWire, Galaxy Zoo, FoldIt and EteRNA in a small category of large-scale resource demanding online citizen cyberscience endeavors where real problem solving is taking place beyond pure data gathering. With this midstream paper, we want to share our experiences from this first year of beta design and player recruitment, and make our reflections and learning curves available. We believe that our findings about the engagement process – although preliminary - contain a series of conclusions pertinent to the design- and engagement processes hidden behind human computation.

We first give a brief introduction to the physics behind the game, and the method of turning its core tenets into a playable game. This paper does not focus on the actual game results and the reader with no interest in physics can safely skip the first part. We then turn to the main focus of the paper, the description of the design considerations from the first year, especially relating to user acquisition and the structural gameplay surrounding the game's core loop. Finally, we present data about participation for the beta year, noting how different properties like recruitment source and physics interest predict tenacity and performance in the game. We conclude by looking closer at our most dedicated "heroes" and discussing future perspectives for human computation and Quantum Moves.

## 2.  THE PHYSICS BEHIND QUANTUM MOVES

Quantum mechanics originated in the beginning of the 20th century when the physicists of the time realized that the known laws of physics were not capable of describing the structure of atoms. Experiments made by Ernest Rutherford showed that the atom had to consist of a positively charged nucleus orbited by negatively charged particles called electrons (Longair, 2003). In 1913, Niels Bohr showed that only certain orbits were allowed, and that the electron could only jump from one obit to another by absorbing or emitting a quantum of light with the correct amount of energy. Even more remarkably the electron was allowed to be in two different orbits simultaneously until a measurement was performed to determine in which of the orbits it was. From this, quantum mechanics evolved through the work of Heisenberg (Born, 1926), de Broglie (1925), Schrödinger (1926) and many others. They showed that atoms should be described by a wave function: A distribution describing the probabilities of measuring the atom in every point in space.

Computer technology has also progressed rapidly over the past decades. Moore's law states that available computer power doubles every 18 months (Moore, 1965), due to the ability to fabricate ever-smaller transistors. However, the miniaturization has a lower limit due to quantum effects. This led to the proposal of the quantum computer (Feynman, 1982), in which the computer bits which are traditionally only allowed to be either 0 or 1 are replaced by quantum bits (qubits) that are allowed to be 0 and 1 simultaneously. A quantum computer holds the potential for huge calculation power since 300 qubits would be capable of representing $2^{300}$ numbers (larger than the total number of atoms in the universe), in contrast to the normal computer where 300 bits can represent just 1 number.

Quantum computers have already been created in a very small system of 7 qubits (Vandersypen, 2001), but they have yet to be implemented in a scalable system capable of outperforming traditional computers. Many proposals for such scalable architectures exist in various systems and in one of them, atoms are contained in an egg-tray like trap made of interfering light beams (Weitenberg, 2011a, Weitenberg, 2011b). The quantum computation is then performed by concatenating a sequence of individual qubit flip operations (Jørgensen, 2014) and two-qubit operations consisting of picking up an atom in one well with a focused tweezer of light and transporting it into contact with another atom somewhere else in the computer (Weitenberg, 2011b). This is non-trivial because any fast movement of the tweezer causes the (probability distribution of the) atom to slosh around and this sloshing will result in an error in the calculation because a sloshing atom contains kinetic energy, which means that it is not in the ground state. The challenge of Quantum Moves consists in finding algorithms describing how the laser in the physical machine should be controlled to move atoms quickly from one location to another without introducing sloshing at the end.

## 2.1   The quantum optimization challenge

In Quantum Moves, players help build a powerful quantum computer by finding ways of moving a simulated atom from one location in the game interface to another, without sloshing. The movement of the atom is guided and constrained by a so-called *potential landscape* spanning the screen (black line in Figure 1). The probability distribution of the atom (green in Figure 1) resembles a liquid, but will slosh and distribute itself in smaller waves at the slightest wrong movement according to the rules of quantum physics. We call the collective shape of the atom at any given instance of time its *state*. Each level describes a unique problem represented by the potential landscape's line combined with pre-specified beginning- and target states. Success is measured by the degree of overlap between the final state of the atom and that of the target area.

A game always consists of controlling the simulated *tweezer* with your computer mouse for a given amount of time (by moving the bottom of an indentation in the potential landscape). Dragging the mouse horizontally changes the position, whereas a vertical move increases or decreases the depth of the trapping indentation (physically realized by turning the power of the laser up or down). A game consists of one complete trajectory of the mouse through a particular level. This can be characterized as the game's "core loop" (Fields, 2014) or (fittingly) "game atom" (Elias, Garfield, & Gutschera, 2012), as it recurs in every game level, variably modified with new obstacles, potential landscapes, goal-states and bonus-points to challenge players or simulate particular physics problems. The levels are organized according to an overall *structural gameplay*, or metagame, where they get unlocked in stepwise fashion and players receive rewards through different symbolic feedback such as high-scores, 1-3 "stars" according to the degree of success, and acquisition of skill- and achievement badges. The entire solution for each game played is stored on our servers for potential future use in our laboratory.

Computer algorithms used to optimize this type of problems typically exist in two variants. In *local algorithms*, an initial trajectory is slightly perturbed in a stepwise fashion, and if a change proves beneficial, further in that same direction. This step – although it sounds simple – can be very sophisticated in state-of-the-art algorithms. Unfortunately, this *deterministic update* often causes it to get stuck in what is called a local maximum. In contrast, algorithms with a *random component* such as genetic optimization algorithms can jump erratically from one solution to the next. This randomness ensures that they will never get stuck and eventually find the best solution – called the global maximum. The disadvantage with this type of optimization is the fact that the steps are random and therefore only beneficial in a small fraction of the times. This makes algorithms with random components exceedingly slow, often in fact so sluggish that in practice they will never converge to the optimal solution. *The aim of Quantum Moves is to combine the best of both worlds in our gamified human quantum optimization: optimization that is rational most of the time, but sometimes makes seemingly random errors or leaps of intuition to rapidly find the sought after solutions.* One example of a gameplay trajectory which requires a certain critical breakthrough (Iacovides, Aczel, Scanlon, & Woods, 2011, 2013) is the level Bring Home Water Fast (see Figure 1). The aim is to fetch an atom from the well on the far right of the screen, and ferry it carefully back to your beginning position on the left. Here, a good solution consists in utilizing the principle of quantum tunneling by bringing the two wells close together without merging them into one. The atom will the tunnel through the classically forbidden intermediate region, and appear in the well created by the moveable tweezer.

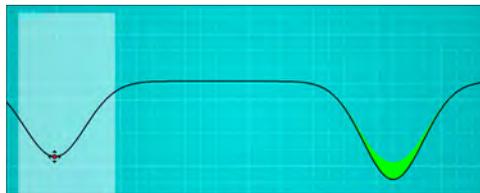

***Figure 1.*** **Illustrations of the level "Bring Home Water Fast"**

Specifically, we compare the players to the computer algorithms in two ways. First, for problems which can eventually be solved by the computer, we compare the score after each optimization to the equivalent high score of players after having played an equal amount of times. As long as the player score is higher than the computer score, the human result really represents the fastest way of getting results at that particular junction. Of course, since the specialty of the computer is to make minute adjustments and improvements, if the particular problem is of a nature that can be solved by the computers, eventually it will overtake the players. In such cases it is an extremely interesting question, to which extent player results can be used as a starting point for a computer optimization that will yield faster convergence rate than the computer optimization alone. Finally, for some of the problems posed the computer fails to find good solutions and it is of course extremely interesting to which extent players can find these solutions.

Although the analysis of the more than 300,000 unique play trajectories generated so far is not complete yet, a pattern seems to emerge that the human optimization is indeed superior to the computer in many problem spaces investigated in all of the three areas discussed above. Even more surprising, it seems that the fraction of the players actually outperforming the computer is quite large. Whether the players succeed by fluke, by building in-game skills, or possibly through a simple theoretical understanding of the quantum physics principles represented in the game is still an open question, which will be the topic of future research.

## 3.   THE NEED FOR SKILLFUL PLAYERS

A central challenge to achieve the optimization described above is that Quantum Moves needs to recruit engageable players and hone their in-game skills over time. Player acquisition is a central part of online game launch strategies. The industry metric *user acquisition cost* (UAC) is an aggregate of advertisement cost, development cost, back-end expenditures and similar prices used to describe how much money a social game developer will spend, on average, to get a new user (Fields, 2014). Developers thus expect to spend quite a bit of time and money to get click-through and installations of their games. This is both an ongoing process of marketing and design, and a focused enterprise to build critical mass at launch. However, while the commercial industry needs to balance this effort and expenditure against each player's average lifetime value (LTV) in dollars and cents, as well as their ability to make the game a social place and recruit friends to join them (*lifetime network value*, LNV), citizen science games need to consider the time and relative price of a different kind of payoff – namely tangible science contribution from each player *and* his network. We can label the average direct player contribution *user science value* (USV).

Player contributions on our sister-game *Galaxy Zoo* have been aggregated into four main clusters, or participation profiles. These reveal that some contribute a lot early, usually on their first login, never to return, while others establish a stable pattern of play over a prolonged period of time (Brasiliero, 2014; Ponciano, Brasiliero, Simpson & Smith, 2014). Each galaxy pattern recognized is a worthwhile addition of data. In more complex games like EyeWire (Robinson, personal communication, 2/21. 2014) and Quantum Moves, however, players need to build a modicum of skill before they can reliably contribute to the core scientific challenges (barring flukes arising from the random factor that is human cognitive and behavioral processing, i.e. Bob, 2009). We could call the average point at which players start doing anything directly useful, the game's *user contribution threshold* (UCT). We have come to call the small percentage of players who tenaciously acquire the necessary skills and persist far beyond the UCT "heroes". Our notion of heroes mirrors that of *whales* from commercial games that rely on free-to-play strategies where only a small group of players are ever really monetized through in-game purchases, premium memberships and the like. While citizen science games with low cognitive complexity can benefit from every minute any player spends, games with high cognitive complexity like Quantum Moves and EyeWire only usually gain direct value from a player if he/she becomes a hero. Players who just visit out of curiosity and then drop out can be called flâneurs, while average participants explore the structural gameplay deeper, and might contribute by fluke. To that calculation we must then add a player's network value, which means that it seems like a viable strategy to make the game fun and attractive for everyone, even if only a few manage to contribute to the core scientific challenge. Generating sustained engagement at less complex levels of game participation may also be a way to gradually build the loyalty and skills needed for a flâneur to transition into a hero. A real concern, which our empirical work aims to address, is, however, if the average hero's preferences and play trajectories diverge substantially from average players.

From a psychological standpoint, it is important to mention the difference between intrinsic and extrinsic motivation, as they apply to participant retention in citizen science projects. There are competing schools of thought on the matter (Deci, 1980; Malone, 1981), but they agree on core tenets: In extrinsic motivation, action is based on outside rewards like money or socially valued praise, or avoidance of unpleasant states like scolding. This entails a tendency to act half-heartedly and cease the behavior when the outside factors dry out (Deci, Koestner, & Ryan, 1999). In a state of intrinsic motivation, on the other hand, the user is driven by desires to participate, explore, learn and master the activity in itself. Extrinsic design-logics are commonly found in vulgar points-badges-leaderboards (PBL) gamification, where mainstream ideas (e.g. Bowser, Preece, & Hansen, 2013; Marczewski, 2012) resemble diluted versions of behaviorist token economy (e.g. Kazdin, 1982; Skinner, 1973) and 20th century utility-based economics brought into question by behavioral economists like Kahneman (2011). Extrinsically informed motivational design places great importance on the magnitude and temporal distribution of rewards, as well as their psychological framing. Intrinsic motivation theories, on the other hand, prescribe design structures that support self-determination via a sense of competence, autonomy and social relatedness (Deci & Ryan, 2008; Rigby & Ryan, 2011), or challenge, fantasy and curiosity (Malone, 1981) plus interpersonal factors like recognition, competition and cooperation (Malone & Lepper, 1987). For an overview, see below Figure 2.

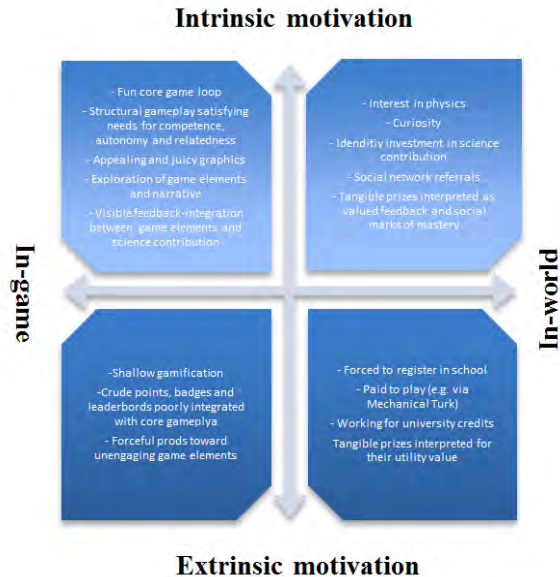

**Figure 2.** *Differences between intrinsic versus extrinsic motivation in in-game versus in-world*

Fogg and Eckles' (2007) "behavior chain for online participation" can be used to conceptualize 3 phases of user involvement. The *Discovery Phase* "Learning about the service" and "Visiting the site. Once they have arrived on the site, users have plenty of opportunities to explore the information available and get the chance "to be educated and influenced". For example, the Quantum Moves site has videos and photos integrated. Also, the website hosts a forum, where visitors can read discussions of registered players and get the chance to be exposed to the game. In the *Superficial Involvement* stage, users are influenced to "Decide to try" and "Get started" with the game. At this stage the structural gameplay will fluently guide them through the tutorial levels and hopefully prompt them to create a profile and validate an email activation link, so they can save their progress. It is only in the final phase, called *True Commitment*, users generate large added value. In Quantum Moves this translates to users making a valuable contribution to either the science, game development, or through their lifetime network value. Such examples range from creation of posts and video materials to creating levels, but contributors' core commitment to Quantum Moves is measured in play counts and scores. This 3-phase process mirrors the conceptual difference between interest (Berlyne, 1954, 1970), motivation (Grant & Dweck, 2003; Jensen & Buckley, 2012; Prestopnik & Crowston, 2011; Raddick et al., 2013; Ryan, Rigby, & Przybylski, 2006) and sustained engagement (Kular, Gatenby, Rees, Soane, & Truss, 2008; Rigby & Ryan, 2011; Skinner, Seddon, & Postlethwaite, 2008). Interest can be understood as the immediate psychological allure of something encountered in the world, usually as a property of perceptual processing (Berlyne, 1954, 1970) exacting a motivational pull of curiosity to investigate. Engagement (Kahn, 1990) includes both motivation, behavioral change, and persistence in an activity.

To acquire engaged players across the beta period, we used four separate recruitment strategies. The early parts of the game (see below) were tested in high-school science classrooms and our own *university lectures*, where large numbers of students were forced to sign up. We also had the opportunity to speak at several high-profile *events*, such as a couple of *public lectures* with over 500 people combined, which we used for an A/B-test of certain game elements (see below). The project has additionally garnered a good deal of attention in traditional *media and online*, which has generated substantial influxes in identifiable jolts. Finally, a fourth group of ongoing clickthrough can be attributed to community efforts and general buzz arising online and in-world, making their origin essentially *unknown*. Since many players cease participating quite quickly (a common pattern in games and citizen science projects alike, but especially prevalent for students forced to participate), we also announced several "featured challenges", where existing players were prompted to come back and knuckle down on particular levels. We awarded extra in-game badges for participation, and offered combinations of books, logo mugs, t-shirts, and even a lab visit with all expenses paid (see below) to top contributors.  We can thus envision a two-dimensional space for motivational devices, with one axis constituted by *in-game* (points, progress, good core gameplay, community, etc.) versus *in-world* (our talks and teaching efforts, prizes, recommendations from friends) (Lieberoth & Roepstorff, 2014; Stevens, Satwicz, & McCarthy, 2007) situation, and the other ranging from *intrinsic* (science participation, challenging gameplay, fun, fascination with physics, etc.) to *extrinsic* (physical prizes, mandatory high school participation) motivational flavors.  Game designs based on intrinsic motivation are thought to satisfy the criteria through gameplay processes alone (the core loop and the structural gameplay surrounding it), while our recruitment strategies in educational settings and prizes in featured challenges can be said to be classically extrinsic. Extrinsic reward has been found to exact a detrimental effect on intrinsic motivations (Deci et al., 1999), but it is worth noticing that points, leaderboard placements and even physical prizes for top-performance may act as intrinsically motivating feedback, social devices and signs of mastery. But people may also already have interests and preferences in the real world, that makes citizen science participation intrinsically motivating, as for instance seen in player data from Galaxy Zoo where the most prevalent motivators were not related to the site's superficial gamification devices, but rather a fascination with outer space or a sense of participating in something greater (Raddick et al., 2013). As such, creating engagement is usually a question of balancing design strategies rather than relying on a single approach, such as vulgar PBL-gamification.

Recruitment activities in-world and online, an engaging in-game core loop, a structural gameplay to frame, structure and motivate the player's continual progression through the levels, as well as an active community where participants get a sense of continually contributing to science, are all central components of the strategy laid out to hopefully realizing the scientific goals of Quantum Moves. In the remaining sections, we describe the game design process as it has unfolded in a sometimes stumbling but always-creative fashion with its many different ideas and theoretical rationales. We then report statistics about users, recruitment and retention, and the attributes that characterize our small crop of heroes so far.

## 4.  DESIGN METHODS

We developed Quantum Moves as the first game under the scienceathome.org umbrella, one of the first citizen science projects in the quantum physics segment. The initiative is developed within our cross disciplinary Centre for Community Driven Research (CODER)[1], where we aim to bridge theoretical and experimental research with online community efforts.

The early development of Quantum Moves, previously known as The Quantum Computer Game, started in December 2011 with the first coding iterations deployed in MATLAB - an algebra program widely used in physics. The decision was based on the program's plotting flexibility and the programming experience available in the student pool.  An early version of the game was ready in February 2012 (see Figure A.1 in the appendix) and subsequently tested in several Danish high schools. This served as an overall proof of concept, yet we noticed that several challenges hindered the game experience, mainly due to MATLAB's limited portability and graphic support. Over the following months, we improved various aspects of the initial prototype. However, a test session with 100 volunteers in the summer of 2012 made it clear that, if we wanted the game to have a large public appeal, we had to abandon MATLAB and rewrite the code in a flexible, end-user accessible programming language. The best solution at that time was Java, as it addressed most of our concerns of end-user accessibility, robustness, flexibility in programming and variety in graphics. A preliminary Java edition came out in early October 2012 (see Figure A.2), and a testing version made available to a small group outside the CODER team in Dec 2012 (see Figure A.3, A.4, A.5). The first, fully-fledged beta version was publicly launched and advertised on the 18th of June 2013 (see Figure A.6). The game was structured around 4 levels: *tutorial*, *Arcade* (a series of games where players could practice) *Scientific* (where we included the games we considered most relevant for our lab research) and *User Space* (a sandbox environment, where players could design their own games or try those created by other users in the community). When the 12 tutorial games were successfully completed, a user was allowed to roam around and try any of the other games found in the *Arcade* and *Scientific* levels.

---

[1] The funding for the CODER center activities are granted from the AU Ideas Programme of the Aarhus University Research Foundation, the Lundbeck Foundation, the John Templeton Foundations, and the Danish Council for Independent Research.

The launch of this beta version was covered in several national media outlets, e.g. National Geographic and videnskab.dk. The media attention played a definite role in making the game known to an audience beyond the usual high school students.  Yet, we experienced several glitches with the university server which slowed down our initial success. A brief follow-up survey sent personally to the top 10 players after launch pointed out that the technical issues were doubled by frustrations with the abrupt increase in difficulty, starting with the tutorial. Also, looking at the data gathered in the days after the release, we noticed, that once players got past the tutorial, they predominately chose to play in the *Scientific* level with insufficient time spent in the *Arcade* level intended to help them acquire necessary core skills. This became a premise for our discussions and, as the team expanded to include a business school graduate and a psychology researcher, we focused on reevaluating the game design. In August 2013, the current Beta version was introduced (Figure A.7). From this point on, we refer to that version in this paper.

Early, we were confronted with two pressing aspects that needed to be addressed: One was redesigning the tutorial to create a lower entry barrier, while also ensuring an effective learning curve. The other was to create a game structure that would enable players to hone the skills needed to perform with high effect in the scientific levels, hopefully turning some into heroes. We started by operationalizing the main physics concepts applied in the game and their equivalent in-game operations into a set of core skills that players would need to acquire: *deceleration* of the atom speed, *tunneling* the atom into the target state and *stabilization* of the atom state. These became our guiding references for the structural redesign of both the tutorial and the advanced levels.  Firstly, we reorganized the tutorial into a set of 7 games, which gradually introduced the main physics elements and core game loop: The atom as a ball, continuing with the atom as a wave, and finishing with the insertion of a static obstacle. To ensure that players had appropriate visual scaffolding and understood the goals of each challenge, we added video animations preceding each level, presenting one possible trajectory along with written hints. The successful completion of the last tutorial level allows players to access the main menu, with the option of playing in separate *skill labs*: Cool, Tunneling and Control.

We decided to create predetermined *paths*, presented in a tree structure (see Figure A.7). By doing so, we aimed to push players to go through the skill training levels before the *scientific levels* where the main contributions to our current citizen science problems are made. Since individual levels differ in difficulty and require refinement of particular in-game operations, each skill lab was divided into a *Bachelor* and *Master* section. The user has to successfully complete 4-6 levels to acquire skill badges on their profile, which would unlock specific new levels. Once the bachelor level was completed, players would gain partial access to both the *Master* section and the *scientific labs*: *QComp* and *Beat AI*. To achieve full scientific access, the Master's would have to be completed at some point.
Beyond this, the structural gameplay tree consisted of a few more nodes: Created as a more theoretically grounded, *FineTune* contains a set of tools that makes it easy for a player to manually change points on the path of his previously played levels, adjust the timing, stretch or shrink the total time or smooth the path. Since the tool functions are not automated, it is ultimately up to the user to decide which parameters are to be changed in order to optimize the path.

The *user-defined space* is open to all players, regardless of skills. It includes *Construction Yard*, a sand box type of environment where users can build their own levels with goals and obstacles and *Playground*, a space where users can play each other's' creations. Compared to the predefined games, these are not connected to the main tree structure, and can be played immediately after completing the tutorial. Therefore, we decided not to set any skills requirements for creating games. Furthermore, since the user games created are primarily supposed to be used as community building drivers, they are neither checked for resolvability, nor included in any of our data analysis processes. Based on the number of plays, the user-created games are ranked on the "The most popular games in the community" leaderboard. The last and newest branch in early 2014 combined predefined user-created games in challenges with the intention to enhance competition and the community feeling. For instance the newest game type is called "Quantum Quests". Here we catered to achievement oriented players (Heeter, Lee, Medler, & Magerko, 2011; Sherry & Lucas, 2006), who crave more traditionally "gamy" elements such as finite Super Mario-like lives. Building on the *iron man* type of game principles, this game consisted of a series of selected games, where a player has to progress as much as possible through the level with only a limited number of attempts, called "lives". It is currently being used as a base for creating more immersive 3D levels using the Unity game development platform.

To create a sense of progress in Quantum Moves with minimal effort, we implemented the simple points-badges-leaderboards (PBL) mechanic recurrent in casual games like *Candy Crush Saga* and shallow marketing gamification. In Quantum Moves, the score is calculated on the basis of various parameters, such as time penalty, overlap with target state, points collected while avoiding the obstacles, and is presented to players in the end screen triggered by the finalization of a game sequence. Based on the obtained score, a player would move up or down on the leaderboard, presented in the lower right corner of the game window (e.g. fig. A4). To mark a player's progress, his score on the leaderboard is shown in relation to similar scores in his range. This is similar to the techniques found on social games, where the scores of a player are presented in relation to others in his/her social network. We made use of two feedback techniques found in gamified applications (Groh, 2012) and mobile games to introduce visual cues that could enhance a player' sense of progression (Deterding, 2012). The first is a bar presented at the top of the game interface, which provides players with real time feedback on their performance. The bar changes its color from red to yellow and finally to green to suggest the effectiveness of a player's trajectory. Mirroring the well-known concept of stars found in games like *Candy Crush Saga*, we also introduced "atoms" as an alternative three-tier grading system reflecting acceptable, good and excellent performance thresholds for each levels, indirectly challenging players to revisit levels and perfecting their collection. The potential downside is that once a game is completed with three stars, players might lose motivation to further optimize his/her score.

The last technique to indicate progress is by acknowledging specific achievements in the form of badges on the player profile (Antin and Churchill, 2011, Denny, 2013). We designed a series of possible badges for two types of achievement: performance and engagement. Depending on performance at various junctions of the game, and cumulative time spent logging and interacting with the game universe, a player could receive a recognition for reaching a specific milestone, "Bachelor degree" or a particular threshold of play counts like "*Quantum Frenzy 350*".

To test the effectiveness of the PBL framework and constrained paths of play on player engagement, we also set up a randomized A/B-test surrounding a major event in August 2013, where around 500 people would attend talks about Quantum Moves and the quantum computer. The literature on gamification is rife with pundits clamoring about the efficacy of badges in motivational design, but there is not much quantitative data to back them up, so we jumped on the opportunity to test whether giving badges to players at various junctions would be a significant motivator for repeated visits and engaged gameplay. At the time, we needed to figure out how to best guide players on their path to the scientific levels, without creating barriers to progress and the feeling of competence. We had conflicting hypotheses about how to build this into structural gameplay: On one hand, the central psychological usability (Norman, 2002) and choice architecture (Thaler & Sunstein, 2009) principle of guiding constraints recommends that a system should have a certain amount of openness, but the basic interface for core operations should be limited to the most preferable options. Locking levels until the necessary skills are accumulated fits this view. On the other hand, we want players to follow their curiosity, maximizing autonomy, motivation, and access to the science levels. In this view, the structural gameplay would not need stringent locks based on skills, but just the tree-like lab structure laid out as an open map, with friendly hints about where to go, if a level proves too difficult. To this end, an "A/B test" was conceived as a 2x2 factorial design, randomly assigning the 500 expected new players to one of four conditions: *Locked* levels or open levels with or without *badges*. Unfortunately, the test was crippled by both time constraints and technical difficulties. Our programmers were working overnight to implement the system to automate A/B-test cell assignment, but also had a lot of more pressing design issues to address before launch. In the end, only a fraction of the badges we had designed were implemented, and players only got weak cues when achieving them. Since we did not have time to test it, the skill system unlocking levels seemed also somewhat counterintuitive. Some old server issues also came back to haunt us on the day, so many invitees were probably unable to log on to the game after creating their account. With these limitations, the usable data were only a fraction of the people assigned to each condition (down to N=1 player in the "Open levels with no badges" cell). The results were statistically insignificant, and will not be reported below. We are awaiting new opportunities to gather this kind of data, which we believe is central to both our own work and adding a much-needed controlled evidence-base to the academic discussion of gamification in general.

As these game design choices were being implemented, we collected data on user acquisition, play trajectories, and the few dedicated players we call heroes from beta launch in 2013 to writing this paper in April 2014. The results are reported in the next section.

## 5.   GAME DESIGN RESULTS: RECRUITING AND LOSING BETA USERS

Between the 10[th] of November 2013 and 10[th] of April 2014[2], there were approximately 6000 users visiting the www.scienceathome.org website, 56.6 % of these being new unique visitors. Average visits last 5 minutes, where users visit an average of 11 pages on the website. IP addresses came from at least 47 countries (see Figure 3); yet, the country tally could be even higher, as we could not place a country of origin to approximately 1000 users. At this point, most registrations originate from Denmark, followed then by Germany and the United States. In the following analysis we present results derived from data collected over a period of 10 months, from the beta version launched on 18[th] of June 2013 until 10[th] of April 2014. The sample relevant for the demographic and human computation data is equal to a cohort of 2511 people.

---

[2] There is no available data for the period 18[th] of July- 9[th] of Nov 2013.

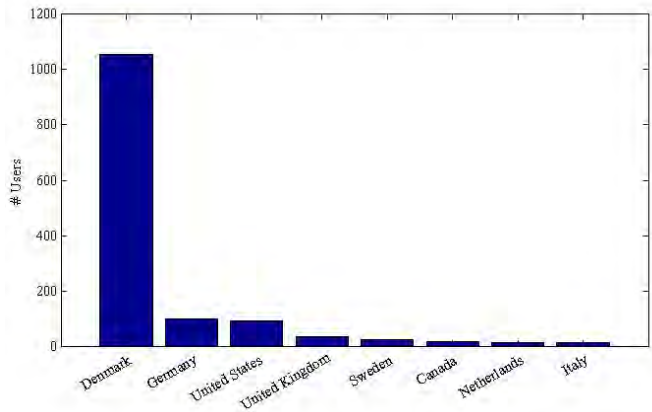

*Figure 3. Player distribution by country.*

60% of all players actually finish the tutorial (see Figure 4), which is an encouraging number. Figure 4 illustrates early drop-off, as people move from superficial involvement to commitment via the 7 mandatory tutorial levels. These data were collected from all users registered between 15th of July 2013 to 10th of March 2014, which is equivalent to a total sample of 1190 players using the newest version of the tutorial.

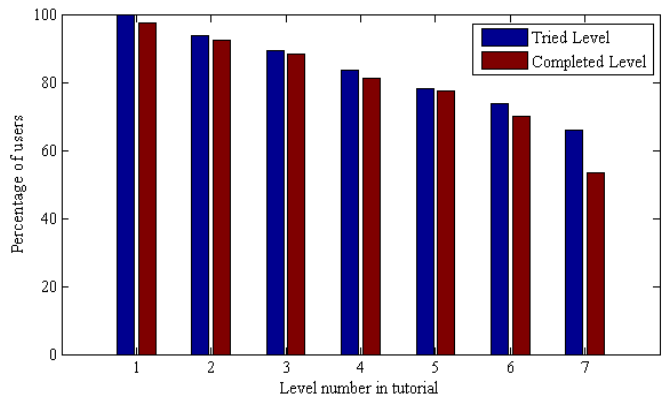

*Figure 4. Tutorial drop-off curve.*

As it can be noticed on Figure 4, the drop-off mainly happens between play during the first six levels (M=98%, 95% CI[97%,99%]). However, the 7th tutorial level display a large discrepancy between the numbers of tried versus completed levels. A one-way ANOVA was performed, to see if the 7th tutorial level (M=81%) belongs to a different normal distribution of completion ratios than the first six levels ($F_{(2, 4)} = 127.12$, $p < 0.001**$). This indicates that the 7th level has a significantly lower completion ratio than the first six tutorial levels, suggesting that the 7th tutorial level, which is the first level that includes the zones of death, which the atom is not allowed to touch, and which constitutes a significant barrier to an otherwise fluid flow of progress, most likely proving too difficult or disrupting competence motivation. An alternative explanation could, however, be that non-committed visitors simply realize that they have now seen all the game has to offer, and either stop because their flanêurs' curiosity has been satisfied, or because they are not sufficiently attracted by the gameplay to engage further.

As detailed at the top of section 4, during the beta period we used four strategies to recruit players. The classifications *Online/Media* and *Voluntary by talks* were based on all registration in a $\pm 3$ days' time span before and after a registered major online or in-world event. The 3 days before the event was included to account for the users that wanted to checkout the game in anticipation of the event. To avoid overlap with *Online/Media*, we attributed the *Voluntary by talks* tag only to the users registered in the same country as where the event took place. The classification of *Forced by talks* was given by a tag added at the point of registration. The results of these recruitment efforts are summarized in figure 5, which displays increases in unique users attributable to one of the 4 main groups.

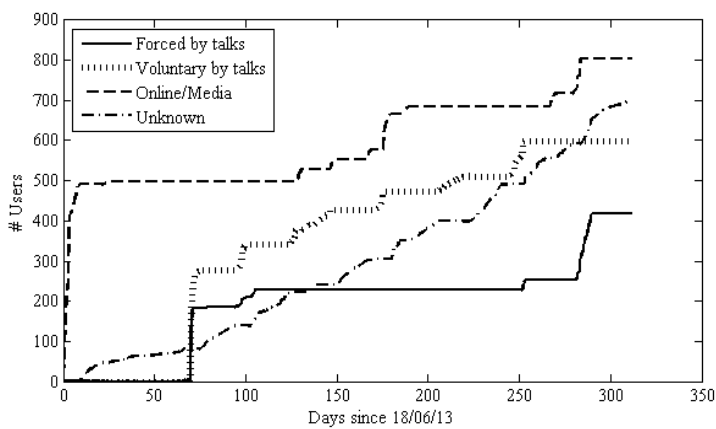

***Figure 5.*** *Users acquired over time in each the recruitment origin groups.*

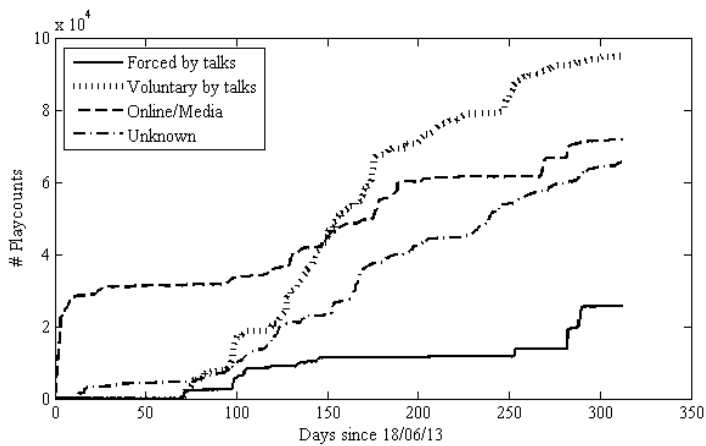

***Figure 6.*** *Play counts for users in each of the recruitment origin groups.*

A large percentage of our registered beta users originate in the *Online/Media* group. The steep initial rise of this curve can be attributed to national media coverage of the game launch in June 2013, while the stepwise increases beginning around day 140 is due to ongoing publicity. From mid-September 2013, we increased our communication efforts by setting up a content strategy for the blog and forum on scienceathome.org, where players ask and answer questions related to the game, report bugs or request new game features. These efforts were complemented by a social media presence, with the launch of a Facebook page and Twitter account, and an increased focus on producing video content for Vimeo. Also, more attention was directed towards a proactive public relations approach, in order to ensure the recognition of the game on established human computation websites and blogs like SciStarter and Citizen Science Centre. Following this, we noticed an increase in the number of other sources mentioning the game or giving positive reviews referring back to our website.

To analyze if any of our data could indicate a predictor for the number of play counts (see Figure 6), a one-way ANOVA test for differences in the number of play counts generated per active day for each player was performed between the four registration origin groups (see Table 1). This showed a significant between-groups difference ($F(3,116)=23.93$, p=<0.001**). In order to discover which groups are different from each other a Tukey-Kramer post-hoc comparison was performed and it shows that both *Forced by talks* and *Voluntary by talks* differed significantly from *Online/Media* (p=0.04* and p<0.001** respectively), *Unknown* (p=0.002** and p=0.002** respectively) and each other (p<0.001**). *Online/Media* and *Unknown* did not differ significantly from each other (p=0.28) suggesting that these two large player groups were qualitatively similar, revealing a pattern where people who actually went out the door in-world on their own accord are highly motivated to play, compared to those who found their way to us online, and especially the poor souls forced by teachers who only played a little.

|  | N users | Mean #plays / day | Lower bound | Upper bound |
|---|---|---|---|---|
| Forced by talk | 417 | 16.95 | 7.58 | 26.33 |
| Voluntary by talk | 374 | 122.5 | 97.62 | 147.4 |
| Online/Media | 750 | 51.29 | 37.21 | 65.37 |
| Unknown | 708 | 69.41 | 50.67 | 88.15 |
|  |  |  |  |  |
| Low physics interest | 192 | 35.27 | 18.53 | 52.01 |
| Middle physics interest | 464 | 32.17 | 25.96 | 38.39 |
| High physics interest | 1242 | 67.2 | 59.27 | 75.14 |
|  |  |  |  |  |
| 0-3 years | 1006 | 70.15 | 57.06 | 83.24 |
| 4-6 years | 463 | 46.58 | 37.74 | 55.41 |
| 7-10 years | 430 | 37.96 | 31.08 | 44.84 |
|  |  |  |  |  |
| Male | 1528 | 73.18 | 62.37 | 84 |
| Female | 371 | 31.36 | 24.23 | 38.49 |

*Table 1. Number of games played per active day.*

A similar procedure was conducted for physics interest, years of physics education beyond 9th grade, and finally between male and female (all summarized in Table 1). The one-way ANOVA shows significant differences for both physics interest ($F_{(2,873)}=11.45$, $p<0.001**$), gender ($F_{(1,582)}=40.37$, $p<0.001**$) and physics education ($F_{(2,873)}=10.87$, $p<0.001**$). Tukey-Kramer post-hoc comparisons show that high interest in physics leads to significantly more play that low and middle interest (both $p<0.001**$), while the latter two do not differ significantly ($p=0.762$). For physics education beyond the 9th grade the post-hoc comparisons show that the very large group with 0-3 years of schooling played much more than those with 4-6 years ($p=0.006**$) and 7-10 years ($p<0.001**$) of additional schooling behind them, which did not differ significantly from each other ($p=0.31$). The game framing of our citizen science project thus seems to speak to people with a notable casual interest in physics, but not a too professional background.

We were also interested in who *performed* well after having completed the tutorial levels, so we calculated the average score on the tutorial levels (the most stable figure, since players may have chosen many different paths after the 7th tutorial level) calculated for the four registration groups, interest, physics education beyond 9th grade, and gender (see Table 2).

|  | N users | Mean point score | Lower bound | Upper bound |
|---|---|---|---|---|
| Forced by talk | 99 | 68.09 | 62.23 | 73.94 |
| Voluntary by talk | 194 | 60.58 | 56.35 | 64.81 |
| Online/Media | 162 | 50.94 | 46.13 | 55.75 |
| Unknown | 338 | 54.48 | 50.39 | 58.57 |
|  |  |  |  |  |
| Low physics interest | 54 | 55.81 | 47.03 | 64.59 |
| Middle physics interest | 125 | 61.11 | 52.68 | 69.54 |
| High physics interest | 495 | 57.07 | 54.32 | 59.83 |
|  |  |  |  |  |
| 0-3 years | 319 | 57.00 | 53.47 | 60.52 |
| 4-6 years | 198 | 56.21 | 51.70 | 60.71 |
| 7-10 years | 157 | 61.10 | 54.35 | 67.86 |
|  |  |  |  |  |
| Male | 585 | 56.49 | 53.66 | 59.32 |
| Female | 89 | 65.80 | 58.68 | 72.93 |

*Table 2. Performance on the tutorial levels*

A one-way ANOVA reveals that the registration groups differed significantly on average score reached on the tutorial levels ($F_{(3,819)}=6.71$, $p<0.001**$). The Tukey-Kramer post-hoc comparisons show that Forced by talks differed significantly from *Online/Media* ($p<0.001**$) and *Unknown* ($p=0.008**$), but not from *Voluntary by talks* ($p= 0.178$). *Voluntary by talks* surprising differ from *Online/Media* ($p=0.035*$) but not *Unknown* ($p=0.132$), which do not differ from each other ($p=0.391$). Interestingly, the *Forced by talks* group performed best, perhaps because they received better in-world instructions than those tackling the tutorial only with in-game cues as scaffolding. But it is also worth noticing, that these players forced by extrinsic means were much less likely to complete the tutorial levels the first place, so the population registered here is likely to reflect only the most tenacious and intrinsically motivated members of the cohort, who might also have engaged fully in the game on their own time.

No significant effect on tutorial scores was found for interest ($F_{(2,671)}=0.76$, p=0.469), nor education $F_{(2,671)}=0.99$, p=0.370), but there was a significant difference between the genders, with females significantly outperforming their male competition ($F_{(1,672)}=5.55$, p=0.019**).

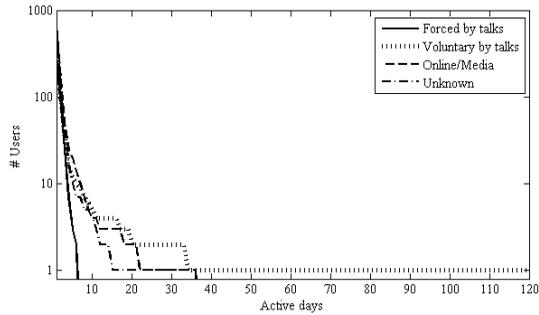

***Figure 7.** Number of users with at least n active days in a specific origin group.*

To assess how Quantum Moves retained players recruited by the four different means, we calculated the number of active days, defined as a day with at least one play count, for each user. A score of 6 can thus mean someone who plays intensely for 6 days on end never to return, or someone who returns a day per months for half a year. A look at the drop-off curves in Figure 7 indicates that 99% of the users drop out within 7 days and never return to the game. A one-way ANOVA comparing the drop-off rates did not show a significant difference between the four registration groups. However, players that stayed for more than 7 active days were all recruited at public events, and subsequently signed up voluntarily. These 19 players are the ones we refer to in our paper as "heroes". Sadly, demographic data is available only for 14 of them. Even though the sample is small, we present a comparative analysis of the heroes versus more casual players in Quantum Moves. In Figure 8, it can be noticed the results derived are based on a sample of 14 "heroes" and 1885 casual users.

## 5.1   Who are the citizen science quantum heroes?

A considerable part of our work has converged on heroes: That is, identifying players who significantly contribute in the science levels based on existing personal attributes like talent, interests and cognitive surplus, and in turn designing structural gameplay to help more casual players cross the UCT through combination of motivation and fluid skill acquisition.

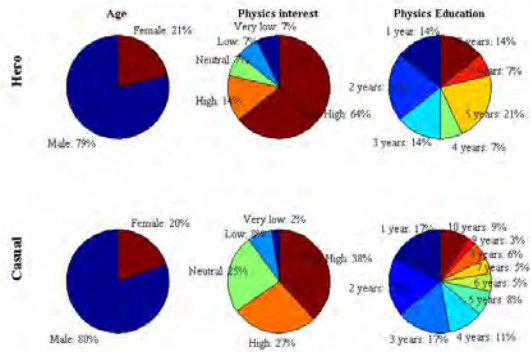

**_Figure 8._** _Characteristics of Heroes (N=14) vs. Casual (N=1885). Left: Gender. Middle: Physics interest. Right panel: Years of physics education beyond 9<sup>th</sup> grade._

A Wilcoxon rank sum test for equal medians was conducted to compare heroes to more casual participants interest in physics (reported on a Likert scale ranging 1-5) (hero mean=4.214 SD=1.311; casual mean=3.907 SD=1.072) and years of education related to physics (e.g. high school science) (hero mean= 4 years, SD=2.418; casual player mean=4.188 years SD= 2.848), revealing no significant predictive power for who becomes a hero on either of these. Firmly believing in the power of mixed methods triangulation, however, we also have some interesting qualitative data to illuminate this. Out of the 14 identified heroes, we had direct contact with three, namely sus, shb and meilby. For privacy reasons, we will refer to them by their Quantum Moves user name. Due to their early contributions, sus and shb were invited as special guests to an offline community event and lab visit on the 6th of November 2013, where they gave brief interviews to help us understand player motivations and promote Quantum Moves. The semi-structured chat was based on 5 questions related predominantly to their motivation to play Quantum Moves and their preferred game elements. Contact to meilby was established via email. His response was among those we received as a reply to the email sent out after the beta version launch in June 2013. In this email, the questions were formulated around the strategies top players used to obtain above average scores in particular games, as well as to give their impressions with regard to the game structure and the newly launched features.

### 5.1.1   Sus

Sus (F, 40 years old) was working as an accountant, based in Copenhagen, Denmark, when we met her. She signed up as a player in Quantum Moves after a public talk on the topic of information security and quantum computers, held by Jacob Sherson at the Experimentarium on 27 -28 August 2013. Since that date, sus had 120 active days and 47570 play counts. This is noteworthy as, according to her own testimony, sus had never played a computer game until she signed up for Quantum Moves. When we asked sus what motivates her to keep playing the game, she mentioned that what motivated her to play Quantum Moves was the "knowledge that the results will be used for real, scientific purposes" (recorded interview with sus, 6th of Nov 2013). Sus' case helps us to understand, that Quantum Moves does not necessarily acquire its heroes from a traditional gamer demographic (although women 30+ are the quickest expanding gamer group, Software Entertainment Association (ESA), 2013), and should take this into account for both recruitment, structural and core game design, and framing purposes.

### 5.1.2    Shb

The drive to help science by playing a computer game was also what Shb (F, 17) mentioned as main motivation to sign up for Quantum Moves. Shb is a Danish high school student and member of the UNF, a youth organization which aims to attract and develop young talents into natural sciences.  Just as Sus, Shb signed up as a player after the public talk at the Experimentarium on 27 -28 August 2013. She has had 19 active days and 5441 play counts. When asked what motivates her to keep playing Quantum Moves, she answered "It is fun to play, among others because I know (that by doing so) I help science" (Danish- English translation of an excerpt from recorded interview with Shb, 6[th] of Nov. 2013). Shb's case suggests that in-world membership in scientific communities of interest may provide a fertile ground for future recruitment.

### 5.1.3    Meilby

Meilby joined the Quantum Moves community early and was part of testing several previous editions of the game before the beta version launch in August 2013. In the selected time span, he had a total number of 17 active days and a 2656 cumulative level plays.

Meilby drives a taxi and has no physics education beyond the standard level obtained by attending high school. His case becomes interesting because he reported being able to replicate, through several iterations, the correct sequences of actions needed to perform an operation known in physics as "quantum tunneling" based on theoretical reasoning rather than trial and error (for full description, see Appendix 9.1). Based on his formal education, he would not have the sufficient knowledge to analytically translate the physics concepts used in the game. Meilby thus exemplifies the contrast between several hypotheses we have about how a hero manages to move to the top of the leaderboards: Through explicit theoretical knowledge about quantum principles (as expected by the physicists in our group, who advocate focusing on explicit semantic knowledge in player skill acquisition), through implicit familiarity and complex processes of predictive coding in the core hand-eye coordination used to play (Friston, 2012; Scott & Dienes, 2008), or some combination of these higher- and lower-order cognitive processes, unique to the human mind/brain/body complex, allowing human players to explore, tinker and strategize their way through difficult and counterintuitive physics problems via more or less conscious trajectories that no computer algorithm would ever fathom.

## 6.  DISCUSSION

In this paper, we have tried to open up our design process for Quantum Moves, and reported significant findings about factors predicting play activity and performance, ranging from gender to recruitment. Notably, we have devoted time to understand the most tenacious and talented kind of players, who we label heroes.

Quantum Moves gets most of its participants online, but our beta data reveal a much more interesting pattern: People who were very interested in physics but had little formal background in the field, and who actually at some point went out the door to meet us, played much more tenaciously than those who clicked their way to www.scienceathome.org online - or were forced by extrinsic means. Thus, genuinely interested amateurs reached in-world seem to be the most fertile ground for recruiting intrinsically motivated citizen cyberscientists to games like Quantum Moves. Having attended talks with real physicists also seems to help people achieve better scores once they start playing. For citizen science, the relationship between in-world and in-game motivation is nontrivial.

As for the actual human computation, the first year of Quantum Moves has been devoted to "calibrating our users" to solve various challenges (first scientific results will be published elsewhere shortly).  We purposely abstained from helping the users beyond teaching them the basic game mechanics, even when we knew of efficient solutions to a problem. This choice provided an unbiased ensemble of solutions, which has two advantages: first it contains solutions we would never have thought of, and second, it provides a measure of what the users can realistically contribute. We use this to compare user learning curves with computer algorithms as they both found solutions for the same level. The score of a user after each try is compared with the score the computer would have after the same number of iterations. This has shown that users are fast at obtaining a good score, but they will never obtain the perfect score, whereas the computer algorithms need a lot more iterations to obtain a good score, but they are capable of finding the solution with a level of precision which yields the perfect score that we need. However, for a level like Bring Home Water Fast the users try out solution-patterns which the computer algorithms would never find by themselves, and they discover solutions which are faster than the computer algorithms' more linear computational trajectory.

Getting to know our player-base and especially the heroes reveals that citizen cyberscience games like Quantum Moves cater to a nontraditional demographic compared to both science and gaming. For instance, female players significantly outscore males after having played the tutorial, even though males play more per day. Our data comparing those forced to play via schoolwork to players who came to us via in-world events or online channels also suggest that users are much more inclined to play if the action springs from curiosity or if players are intrinsically motivated to contribute to science, supporting earlier findings by Raddick and colleagues (2013) as well as the central tenets of self-determination theory. What is less clear, however, is what game elements help engage and retain players once they have been recruited, and begin to move from superficial to true commitment.

While Quantum Moves is unique compared to other citizen science games in having an engaging and challenging core game loop that *by itself* lives up to prominent definitions of (casual) games (Juul, 2005; Salen & Zimmerman, 2004), we also expect that a well-designed structural gameplay (sometimes called metagame) is central to frame, structure and motivate the play experience, both helping and goading players to move from level to level along appropriate learning curves balanced between boredom and anxiety. In the end, we accept that the high level of cognitive complexity in Quantum Moves compared to citizen science games that rely on players as simple pattern-hunters or "mules" carrying data gathering devices into the real world , means that we will lose players at a higher rate due to the difficulty, but we believe that it is a viable strategy to work towards a game that is fun and learnable for everyone: On one hand to capitalize on each player's *lifetime network value*, and on the other in the hope of helping players hone the three skills central to succeed at the *scientific levels*, moving them beyond the user contribution threshold.

We were vague about the difference between a hero and a player who crosses the UCT. This is because we view hero-status as a significant individual property combining sustained engagement and a high level of skill, while crossing the UCT can happen to anyone by a fluke or one good gameplay. In this sense, we conceive of our heroes less as loincloth clad Conan-types who individually topple empires and more alike the WW1 trench-heroes who made up the bulge with grit, skill, and intrinsic determination, and together make a difference worth reverence. However, even though many players may be truly engaged and come back for several days, we still need deeper analysis of the scientific results to tell how many actually manage to cross the UCT by accident and/or by attaining hero status – that is, who actually manages to contribute to the quantum computer, as our beta population has mainly guided us in designing an effective game interface, and test simple hybrid optimization against only specific algorithms.

The limits to the current study are manifold, as we have compiled this contribution as an open midstream report rather than waiting for the best possible data years into the future. Centrally, the N of active players, and especially heroes, is limited, and mainly stem from a Danish context. As such, new patterns are likely to emerge when we establish an even wider international profile. Also, the many ANOVAs run are almost certain to show some kind of statistically significant results, even though they may not be practically significant. MANCOVAs would have been preferable, but since the population for each test differed slightly, this is the right kind of statistical test available at this point. Further analysis and data gathering is clearly needed. In this instance, our A/B test was unable to show a statistically significant role for challenging constraints, openness to explore or the infamous. The lesson learned is, that large-scale tests should only be conducted after a satisfactory alpha-test of the gameplay in each cell, and ideally supported by using qualitative measures of player engagement and interaction-patterns to allow triangulation of causal mechanisms shaping gameplay trajectories. The games industry might not have time for these kind of detailed studies with most casual games being developed by small expert teams in less than 6 months, but as scientists, we should. So much the wiser, we still hope to one day lead the citizen cyberscience games community in methodologically sound testing of design hypotheses about recruitment, engagement and different kinds of motivation.

The relationship between curiosity on one hand, and intrinsic versus extrinsic motivation on the other, goes some way toward explaining the steep and permanent drop-off in player engagement as they make their way through the tutorial, and is especially seen after Quantum Moves teaching and recruitment events, where people may have been forced to register and thus driven primarily by external factors. A steep drop-off curve is in no way unusual for online games (Fields, 2014), so a full 60% tutorial playthrough suggests that the levels encountered early in the behavior chain actually keep players interested for a good while. Future design perspectives include building better just-in-time feedback into the core game loop, making it more 'juicy' and helping players understand exactly what is going well or wrong at any moment. After a year of stalling due to programming pool limitations, we also finally have the capacity to integrate the game with social media such as Facebook, which will add an entirely new social dimension to both gameplay and recruitment. Finally, we have yet to narratively structure the structural gameplay and implement truly juicy graphics. As our community manager noted, we do not need to become the next Candy Crush Saga in terms of juiciness and engagement – just the Candy Crush Saga of physics.

We are now in a position to deploy more advanced psychological methods in mapping player trajectories and trying to predict who might be born heroes and who can become ones, which should significantly inform our future design of the structural gameplay. Our Centre's unique cross-disciplinary nature allows us access to eye-tracking and other tools that will soon help us test interaction design, and conflicting hypotheses about the usefulness of *explicit* understanding of the physics issues (as expressed my Meilby) versus *implicit* learning of the hand-eye coordination and establishment of predictive coding (Bubic, von Cramon, & Schubotz, 2010; Friston, 2012) to account for the counterintuitive movement of the 'liquid' wave. As another exiting step we will soon be able to report the first comprehensive psychometrics battery collated especially for citizen science gaming, in collaboration with other major players. The revolutionary bit will, however, not be the our data collection, but the analysis of complex relationships between hosts of in-game and in-world variables via our next game: A title both for teaching statistics and engaging our community in analyzing data with advanced techniques like path analysis, that require intelligent human causal propositions and critical evaluations of other user's solutions – something no algorithm can ever do.

## 7.  CONCLUSIONS AND FUTURE PERSPECTIVES

Human optimization is superior to the computer in many instances, but we have yet to understand the exact cognitive nature of these solutions as they apply to the varying problem spaces in Quantum Moves. This, along with more work on motivation, will be the subject of extensive future research.

Contrary to commercial casual games, we cannot change or abandon our core game loop, which represents the real quantum optimization processes needed to operate a scalable quantum computer. We can, however, change the look, feel, usability, and structural gameplay surrounding our game to frame, structure and motivate more engaging play trajectories, and ensure a sufficient learning curve to help people cross our high science contribution threshold. We have opened up that messy and experimental design process in this inaugural issue, hoping that others will be able to learn from our experiences. We encourage other researchers to publish future Human Computation pieces along the same lines.

## 8. ACKNOWLEDGEMENTS

[1] The funding for the CODER center activities are granted from the AU Ideas Programme of the Aarhus University Research Foundation, the Lundbeck Foundation, the John Templeton Foundations, and the Danish Council for Independent Research

# 8.  REFERENCES

Antin, J., & Churchill, E. F. (2011). Badges in social media: A social psychological perspective. In *CHI 2011 Gamification Workshop Proceedings, 7 -12 May 2011*, Vancouver, BC, Canada.

Berlyne, D. E. (1954). A theory of human curiosity. *British Journal of Psychology, 45*, 180–191.

Berlyne, D. E. (1970). Novelty, complexity, and hedonic value. *Perception and Psychophysics, 8*, 279–286.

Bob, P. (2009). Chaos, cognition and the disordered brain. *ANS: The Journal for Neurocognitive Research, 50*, 114–117. Retrieved from http://w.activitas.org/index.php/nervosa/article/viewArticle/14

Bowser, A., Preece, J., & Hansen, D. (2013). Gamifying Citizen Science : Lessons and Future Directions. *Proceedings of CHI 2013 Workshop Designing Gamification: Creating Gameful and Playful Experiences,* 5

Brasiliero, F. V. (2014). Volunteers engagement profiles in volunteer thinking systems. In *Presentation at Citzen Cyberscience Summit 2014, 20-22. February 2014.* London, UK.

Bubic, A., von Cramon, D. Y., & Schubotz, R. I. (2010). Prediction, cognition and the brain. *Frontiers in Human Neuroscience, 4*(March), 25. doi:10.3389/fnhum.2010.00025

Deci, E. L. (1980). *The psychology of self-determination.* Lexington, MA: D. C. Heath (Lexington Books).

Deci, E. L., Koestner, R., & Ryan, R. M. (1999). A meta-analytic review of experiments examining the effects of extrinsic rewards on intrinsic motivation. *Psychological Bulletin, 125*(6), 627–68; discussion 692–700. Retrieved from http://www.ncbi.nlm.nih.gov/pubmed/10589297

Deci, E. L., & Ryan, R. M. (2008). Self-determination theory: A macrotheory of human motivation, development, and health. *Canadian Psychology, 49*(3), 182–185.

Denny, P. (2013) The effect of virtual achievements on student engagement. In Proceedings of the SIGCHI Conference on Human Factors in Computing Systems (CHI '13). ACM, New York, NY, USA

Deterding, S. (2012) Gamification: designing for motivation, *Magazine Interactions 19,* 14

Edelman, G. M., & Tonini, G. (2000). *A universe of consciousness. New York: Basic.* London: Penguin.

Elias, G. S., Garfield, R. S., & Gutschera, K. R. (2012). *Characteristics of Games.* Cambridge MA: Mit Press.

Fields, T. (2014). *Mobile & Social Game Design: Monetization Methods and Mechanics* (2nd ed.). Boca Ranton, FL: CRC press - Taylor & Francis. R

Fogg, B.J., & Eckles, D., (2007), *Mobile Persuasion: 20 Perspectives on the Future of Behavior Change.* Stanford Captology Media

Friston, K. (2012). Predictive coding, precision and synchrony. *Cognitive Neuroscience.* doi:10.1080/17588928.2012.691277

Grant, H., & Dweck, C. S. (2003). Clarifying achievement goals and their impact. *Journal of Personality and Social Psychology, 85*(3), 541–53. doi:10.1037/0022-3514.85.3.541

Groh, F. (2012) Gamification: State of the Art Definition and Utilization, *Proceedings of 4th Seminar on Research Trends in Media Informatics* , 39

Heeter, C., Lee, Y., Medler, B., & Magerko, B. (2011). Beyond player types: gaming achievement goal. In, Sandbox '11: Proceedings of the 2011 ACM SIGGRAPH *Symposium on Video Games* (pp. 43–48).

Iacovides, I., Aczel, J., Scanlon, E., & Woods, W. (2011). What can breakdowns and breakthroughs tell us about learning and involvement experienced during game-play? In *5th European Conference on Games Based Learning, 20-21 October 2011.* Athens, Greece. R

Iacovides, I., Aczel, J., Scanlon, E., & Woods, W. (2013). Making Sense of Game-Play : How Can We Examine Learning and Involvement ? *Transactions of the Digital Games Research Association*, *1*(1), 1–17. Retrieved from http://todigra.org/index.php/todigra/article/view/6

Jensen, E., & Buckley, N. (2012). Why people attend science festivals: Interests, motivations and self-reported benefits of public engagement with research. *Public Understanding of Science, Bristol, England*

Juul, J. (2005). *Half-real: Video Games between Real Rules and Fictional Worlds*. Cambridge MA: MIT press.

Kahn, W. (1990). Psychological Conditions of Personal Engagement and Disengagement at Work. *Academy of Management Journal*, *33*(4), 692–724.

Kahneman, D. (2011). *Thinking, Fast and Slow*. London: MacMillan.

Kazdin, A. E. (1982). The token economy: a decade later. *Journal of Applied Behavior Analysis*, *15*, 431–445. doi:10.1901/jaba.1982.15-431

Kular, S., Gatenby, M., Rees, C., Soane, E., & Truss, K. (2008). *Employee Engagement : A Literature Review*. Kingston.

Lee, J., W. Kladwang, M. Lee, D. Cantu, M. Azizyan, H. Kim, A. Limpaecher, S. Yoon, A. Treuille, R. Das,  and EteRNA Participants (2014)  RNA design rules from a massive open laboratory, *PNAS* 2014 111 (6) 2122-2127

Lieberoth, A., Marin, C. M. & Møller, M. (in review). Deep and shallow gamification: the thin evidence for effects in marketing and forgotten powers of good games. In J. Martí-Parreño, C. Ruiz-Mafé, & L. L. Scribner (Eds.), *Engaging Consumers through Branded Entertainment and Convergent Media*. IGI global.

Lieberoth, A., & Roepstorff, A. (2014). Mixed methods in games research – playing on strengths and countering weaknesses. In P. Lankoski & S. Björk (Eds.), *[untitled game studies methods book] - in press*. Pittsburgh, PA: ETC Press.

Malone, T. (1981). Toward a theory of intrinsically motivating instruction*. *Cognitive Science*, *4(3)*, 333-369

Malone, T., & Lepper, M. (1987). Making Learning Fun - A Taxonomy Of Intrinsic Motivations *Aptitude, learning, and instruction, 3, 223-253*

Marczewski, A. (2012). *Gamification – a simple introduction. Tips, advice and thoughts on gamification*. Amazon.com: Self-published for Kindle.

Norman, D. A. (2002). *The Design of Everyday Things* (2002nd ed.). New York: Basic Books.

Ponciano, L., Brasileiro, F., Simpson R., and Smith A. (2014), Volunteers' Engagement in Human Computation Astronomy Projects, Computing in Science & Engineering 99, 1

Prestopnik, N., & Crowston, K. (2011). Gaming for (citizen) science: exploring motivation and data quality in the context of crowdsourced science through the design and evaluation of a social-computational. e-Science Workshops (eScienceW), 2011 IEEE Seventh International Conference, Stockholm

Raddick, M. J., Bracey, G., Gay, P. L., Lintott, C. J., Cardamone, C., Murray, P., & Vandenberg, J. (2013). Galaxy Zoo: Motivations of Citizen Scientists. *Astronomy Education Review*, *12*, 010106.

Rigby, C. S., & Ryan, R. M. (2011). *Glued to games - How Video Games Draw Us In and Hold Us Spellbound*. Praeger.

Ryan, R. M., Rigby, C. S., & Przybylski, A. (2006). The Motivational Pull of Video Games: A Self-Determination Theory Approach. *Motivation and Emotion*, *30*(4), 344–360. doi:10.1007/s11031-006-9051-8

Salen, K., & Zimmerman, E. (2004). *Rules of Play: Game Design Fundamentals*. *Leonardo* (Vol. 37, p. 670). doi:10.1093/intimm/dxs150

Scott, R. B., & Dienes, Z. (2008). The Conscious, the Unconscious, and Familiarity. *Journal of Experimental Psychology: Learning, Memory, and Cognition*, *34*(5), 1264–1288.

Sherry, J., & Lucas, K. (2006). Video game uses and gratifications as predictors of use and game preference.In P. Vorderer & J. Bryant (eds.) Playin computer fames: Motives, responses and consequences (pp. 213-244) Mawah, NJ: Lawrence Erlbaum

Skinner, B. F. (1973). The free and happy student. *Phi Delta Kappan*, *55*(1), 13–16.

Skinner, N. C., Seddon, K., & Postlethwaite, K. C. (2008). Creating a model to examine motivation for sustained engagement in online communities. *Education and Information Technologies*. doi:10.1007/s10639-007-9048-2

Software Entertainment Association (ESA). (2013). *2013 Essential Facts About the Computer and Video Game Industry*.

Stevens, R., Satwicz, T., & McCarthy, L. (2007). In-Game , In-Room , In-World : Reconnecting Video Game Play to the Rest of Kids ' Lives. In *The ecology of games* (pp. 41–66). doi:10.1162/dmal.9780262693646.041

Thaler, R. H., & Sunstein, C. R. (2009). *Nudge: Improving Decisions About Health, Wealth, and Happiness*. London: Penguin Group US.

Websites
ScienceatHome, Home. (n.d.).ScienceatHome, Home. Retrieved April 23, 2014, from http://www.scienceathome.org
Quantum Moves, Home. (n.d.).ScienceatHome, Home. Retrieved April 23, 2014, from http://www.scienceathome.org

## 9.  APPENDIX:

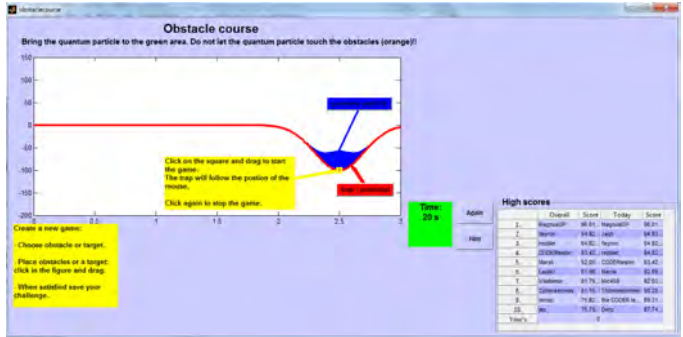

*Figure A.1. The front menu of The Quantum Computer Game in MATLAB version*

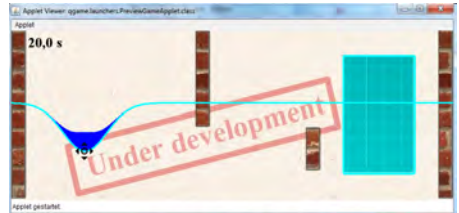

*Figure A.2. One game level in the Quantum Computer Game*

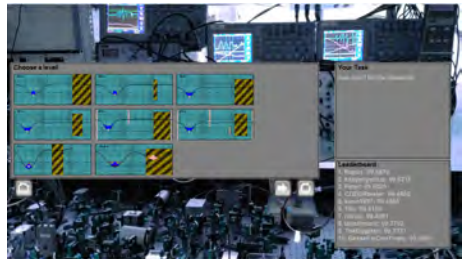

*Figure A.3. Front menu of the first Java version of The Quantum Computer Game*

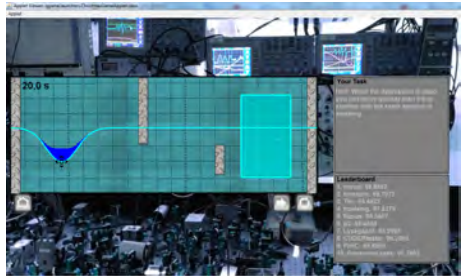

*Figure A.4. Example of game in the first beta version of The Quantum Computer Game*

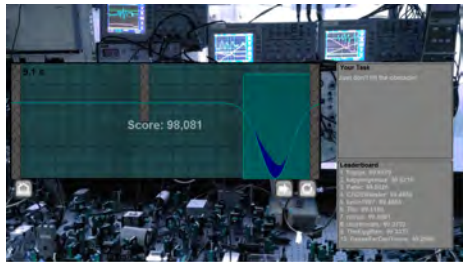

*Figure A.5. Example of feedback window once a game was completed in the first Beta version of the Quantum Computer Game*

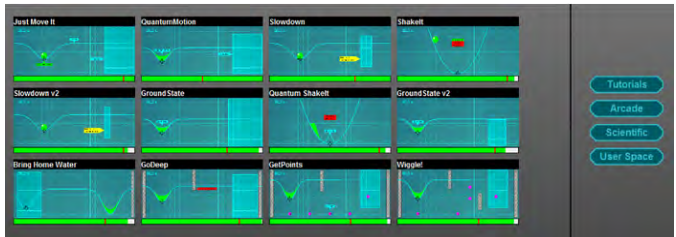

*Figure A.6. In-game front menu of the beta version launched in June 2013*

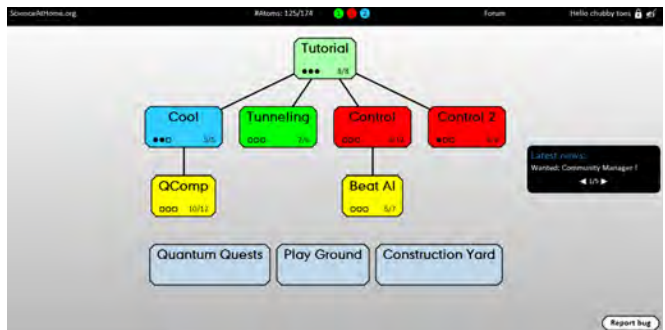

*Figure A.7. Tree structure of Quantum Moves. The in-game front menu of the current Beta version*

## 9.1   Interview with meilby

"Merge Merge was a pain (…) and since I did not really figured out how to get a single atom to the excited state I gave up kind of fast. I got the best score I think it was 3,4, and I was a little mad about being so bad at the challenge because the score went to 692 as max, and 3.4 is just not even near it.

Well I had some time off, and I tried again a, or some days after. With no luck.

Then I said to myself, since I was not doing any progress, and did not know how is should get the first atom to the excited state, I started doing the Excited challenge again, and found the only way I could do it.

Since the background was with lines I started using them, I started raising the curve with the first atom to just above the second atom (according to the lines) and figured I was letting then merge too fast because the first atom was what I would say too excited, and not going to relax. Then I had to be patient, when they were merging, and with some patience letting it flow in itself when they got near each other. The first times it did not work out, instead of the $1^{st}$ atom splitting into 2 it spilt into 3, and I knew I had it I just needed more tries.

Then it happened after some cursing it split into two, and I got a score 16 points, which I was chocked about, I was like WTF, I just did it, and I am getting nothing!

I think it is too hard to do, and the fact that you cant do it fast is giving me stress, the fact I have to wait all 20 seconds every time to make it happen is a reason why it's not so funny to do. (I guess I played too much, and hate the fact I can't get it done right, pisses me off J)

Then I started finetune my score 16 point, I knew already it was done right because it spilt into 2 and the $2^{nd}$ atom was really relaxed.

It took me some time to be familiar with the finetune part, but after spending some time in there, I finally got something, by zoom and keep moving small parts and watching the score I got to 460 points. I used the smooth tool early on, because it was flickering the line in the mittle graph, and after I had dragged the line around and around, it started to flicker (jeg mener linjen bliver zig zagget), I believe it is not good for the score, but i'm not able to remove it, I tried the locking tool to lock the line around the part and the use the smooth tool on the not locked line but it fucked it all up. I've added some pictures to show you what happened."